\documentclass[aps,prl,twocolumn,showpacs,superscriptaddress]{revtex4-1}

\usepackage[dvipdfm]{graphicx}
\usepackage[dvipdfm]{hyperref}
\usepackage{bm}

\newcommand{\unit}[1]{~\mathrm{#1}}
\newcommand{\shiki}[1]{Eq.~(\ref{#1})}
\newcommand{\zu}[1]{Fig.~{\ref{#1}}}
\newcommand{\hyou}[1]{Table~{\ref{#1}}}
\newcommand{\refs}[1]{Refs.~{\cite{#1}}}
\def\tr{\mathrm{tr}}
\def\tk{\tilde{\kappa}}
\newcommand{\To}{T_{\oplus}}
\newcommand{\Omegao}{\Omega_{\oplus}}
\newcommand{\omegao}{\omega_{\oplus}}
\newcommand{\betao}{\beta_{\oplus}}
\newcommand{\rot}{{\rm rot}}
\newcommand{\cx}{{\rm c}}
\newcommand{\sx}{{\rm s}}

\begin{document}

\title{New Limit on Lorentz Violation Using a Double-Pass Optical Ring Cavity}

\author{Yuta Michimura}
  \email{michimura@granite.phys.s.u-tokyo.ac.jp}
  \affiliation{Department of Physics, University of Tokyo, Bunkyo, Tokyo 113-0033, Japan}
\author{Nobuyuki Matsumoto}
  \affiliation{Department of Physics, University of Tokyo, Bunkyo, Tokyo 113-0033, Japan}
\author{Noriaki Ohmae}
  \affiliation{Department of Applied Physics, University of Tokyo, Bunkyo, Tokyo 113-8656, Japan}
\author{\\Wataru Kokuyama}
  \affiliation{National Metrology Institute of Japan, National Institute of Advanced Industrial Science and Technology (AIST), Tsukuba, Ibaraki 305-8563, Japan}
\author{Yoichi Aso}
  \affiliation{Department of Physics, University of Tokyo, Bunkyo, Tokyo 113-0033, Japan}
\author{Masaki Ando}
  \affiliation{National Astronomical Observatory of Japan, Mitaka, Tokyo 181-8588, Japan}
\author{Kimio Tsubono}
  \affiliation{Department of Physics, University of Tokyo, Bunkyo, Tokyo 113-0033, Japan}

\date{\today}

\begin{abstract}
A search for Lorentz violation in electrodynamics was performed by measuring the resonant frequency difference between two counterpropagating directions of an optical ring cavity. Our cavity contains a dielectric element, which makes our cavity sensitive to the violation. The laser frequency is stabilized to the counterclockwise resonance of the cavity, and the transmitted light is reflected back into the cavity for resonant frequency comparison with the clockwise resonance. This double-pass configuration enables a null experiment and gives high common mode rejection of environmental disturbances. We found no evidence for odd-parity anisotropy at the level of $\delta c /c \lesssim 10^{-14}$. Within the framework of the standard model extension, our result put more than 5 times better limits on three odd-parity parameters $\tk^{JK}_{o+}$ and a 12 times better limit on the scalar parameter $\tk_{\tr}$ compared with the previous best limits.
\end{abstract}


\maketitle

{\it Introduction.}---The formulation of the theory of special relativity revealed Lorentz invariance as the fundamental symmetry of the Universe. Since Einstein's first paper over 100 years ago~\cite{Einstein}, a wide variety of experimental tests have been carried out, but no violation has been found~\cite{*[{For a review, see, for example: }] [{}] Mattingly}. As a consequence, Lorentz invariance underlies all the theories of fundamental interactions such as the standard model of particle physics and general relativity. However, theoretical works towards the unification of fundamental interactions suggest Lorentz violation at some level~\cite{LorentzViolation,SME}. Additionally, the observed anisotropy of the cosmic microwave background suggests a possible preferred frame in the Universe. These suggestions have motivated the experimental search for violations with increasing precision.\par

In order to compare the precision of various experimental tests of Lorentz invariance, the theoretical framework of the minimal standard model extension (SME)~\cite{SME} has been widely used. One of the most traditional and direct ways to test special relativity is to test the constancy of the speed of light $c$, or the Lorentz invariance in photons. SME parametrizes Lorentz violation in the photon sector with 19 independent parameters, which consist of 10 parameters ($\tk^{JK}_{e+}$ and $\tk^{JK}_{o-}$) representing vacuum birefringence, 5 parameters ($\tk^{JK}_{e-}$) representing directional dependence in the speed of light, 3 parameters ($\tk^{JK}_{o+}$) representing the relative difference between the speed of light propagating in opposite directions, and 1 parameter ($\tk_{\tr}$) representing the isotropic shift in the speed of light. Here, $J$ and $K$ run from $X$ to $Z$, which represent the spatial coordinate axes of the Sun-centered celestial equatorial frame (SCCEF)~\cite{Birefringent}.  \par

There are tight constraints on the birefringent parameters $\tk^{JK}_{e+}$ and $\tk^{JK}_{o-}$ at the level of $10^{-32}$~\cite{Birefringent}, which were set by polarization measurements of light from cosmologically distant sources. Modern versions of Michelson-Morley (MM) experiments using two orthogonal optical cavities put upper limits on the even-parity parameters $\tk^{JK}_{e-}$ at the $10^{-17}$ level and the odd-parity parameters $\tk^{JK}_{o+}$ at the $10^{-13}$ level~\cite{Eisele, Herrmann}. The degradation in the sensitivity to $\tk^{JK}_{o+}$ compared with that to $\tk^{JK}_{e-}$ comes from the fact that MM-type experiments use even-parity symmetric optical cavities. The degradation factor is given by the orbital velocity of Earth: $\betao = v_{\oplus}/c \simeq 10^{-4}$. The sensitivity to the scalar parameter $\tk_{\tr}$ is further degraded by a factor of $\betao^2$, and the best limit set by an MM-type experiment is at the $10^{-9}$ level~\cite{Hohensee}.\par

Only a few odd-parity tests on the constancy of the speed of light have been carried out so far. Interferometers or cavities will have direct sensitivity to odd-parity parameters only if they are made asymmetric. Although it was not analyzed within the framework of SME, the first test using an asymmetric interferometer was done by Trimmer~\cite{Trimmer}. Recently, improving the sensitivity by changing the Trimmer-type triangular Sagnac interferometer to a ring cavity was proposed~\cite{Tobar2005,Exirifard} and demonstrated \cite{Baynes}. They searched for a nonzero resonant frequency difference between two counterpropagating directions of an asymmetric optical ring cavity and put upper limits on $\tk^{XZ}_{o+}$ at the $10^{-13}$ level and $\tk_{\tr}$ at the $10^{-9}$ level.\par

In this type of experiment, we improved the constraints on $\tk^{JK}_{o+}$ by a factor of more than 5 and $\tk_{\tr}$ by a factor of 12, by making use of a double-pass configuration. Although tighter constraints on these parameters have been derived from astrophysical observations~\cite{Klinkhamer} or particle accelerators~\cite{Hohensee2009,Altschul}, they are indirect constraints deduced from theoretical assumptions~\cite{DataTables}. Our new constraints are the best constraints ever obtained from direct measurement.

\begin{figure}
\begin{center}
\includegraphics[scale=1.25]{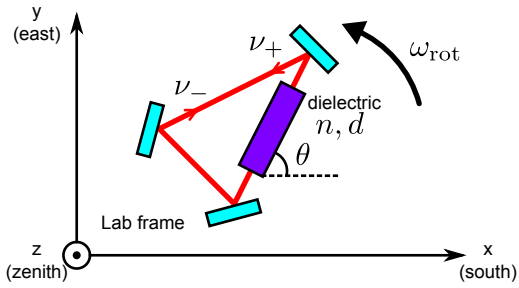}
\end{center}
\caption{\label{ringcavity} Optical ring cavity containing a dielectric. The coordinate axes of the standard laboratory frame are also shown. The resonant frequencies of clockwise ($\nu_{-}$) and counterclockwise ($\nu_{+}$) directions are different if the Lorentz invariance in electrodynamics is violated.}
\end{figure}

{\it Asymmetric optical ring cavity.}---A ring cavity will have direct sensitivity to the odd-parity Lorentz violation if the refractive index changes asymmetrically through its path. Figure~\ref{ringcavity} shows a triangular ring cavity with a piece of dielectric material placed along one side of the triangle. The general expression and its derivation for the resonant frequency shift ($\delta \nu$) due to the Lorentz violation can be found in \refs{Birefringent,Tobar2005}. For the cavity shown in \zu{ringcavity}, this shift is calculated as
\begin{eqnarray}
  \frac{\delta \nu_{\pm}}{\nu} &\equiv& \frac{\nu_{\pm} - \nu}{\nu} = \mp M R^{iI} k^{I} \cdot \left(
\begin{array}{c}
  \cos{\theta} \\
  \sin{\theta} \\
  0 \\
\end{array}
\right) , \label{dnu/nu} \\
  M &\equiv& \frac{(n-1)d}{L+(n^2-1)d} , \label{eqM}
\end{eqnarray}
where $L$ is the round-trip length of the cavity, $d$ is the path length inside the dielectric, $n$ is the refractive index of the dielectric, and $\theta$ is the angle between the dielectric and the laboratory frame $x$ axis. Here, we assumed the SME birefringent parameters and even-parity parameters to be zero and the relative magnetic permeability of the dielectric to be $\mu_{\rm r} \simeq 1$. $R^{iI}$ is the rotation from the SCCEF to the standard laboratory frame~\cite{Birefringent}, and $k^{I}$ is
\begin{equation}
 k^{I} \equiv \frac{1}{2} \varepsilon^{IJK} \tk_{o+}^{JK} + 2 \betao^{I} \tk_{\tr} ,
\end{equation}
where $\varepsilon^{IJK}$ is the Levi-Civita symbol.\par
As indicated by \shiki{dnu/nu}, the frequency shift is modulated by rotating the cavity, and the photonic SME parameters can be extracted by demodulating the frequency shift signal. As the signs of the frequency shift are opposite between the clockwise ($\nu_{-}$) and counterclockwise ($\nu_{+}$) directions, measuring the resonant frequency difference between two counterpropagating directions gives us the Lorentz violation signal. Also, this differential measurement is highly insensitive to environmental disturbances because the effects of cavity length fluctuations are common to both resonances.\par

\begin{figure}
\begin{center}
\includegraphics[scale=1.25]{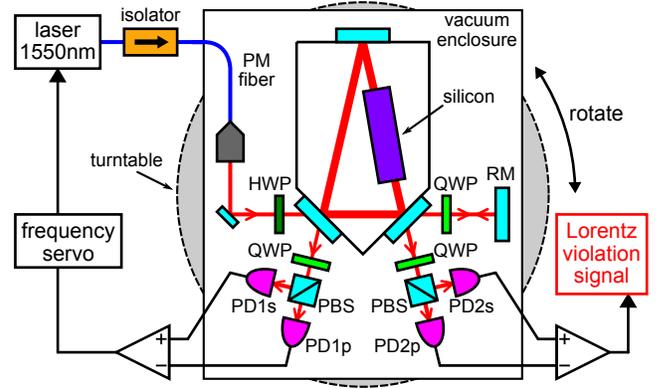}
\end{center}
\caption{\label{opticalsetup} Schematic of the experimental setup; PD = photodetector, PBS = polarizing beam splitter, HWP = half-wave plate, QWP = quarter-wave plate, RM = reflection mirror. The differential of the PD1s and PD1p outputs is the error signal for the laser frequency servo and the differential of the PD2s and PD2p outputs is the Lorentz violation signal.}
\end{figure}

\begin{figure*}
\begin{center}
\includegraphics[scale=0.605]{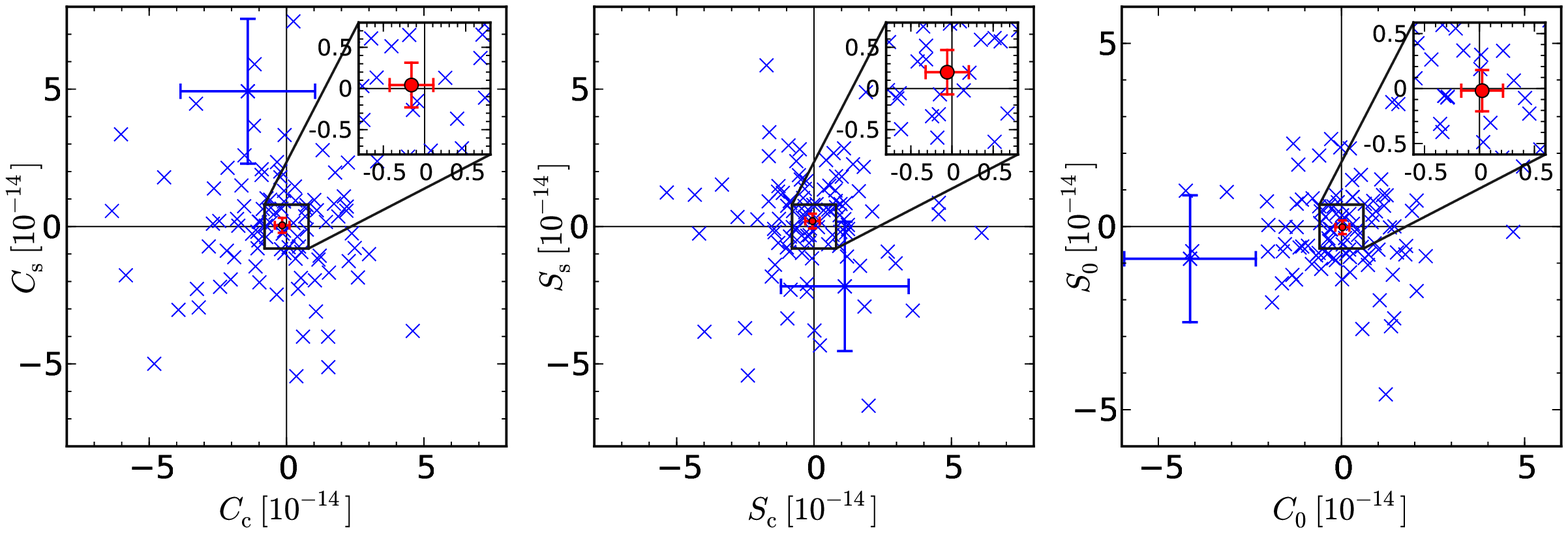}
\end{center}
\caption{\label{phasespace} Sidereal modulation amplitudes determined from 1 day sets of $\mathcal{C}$ and $\mathcal{S}$. For clarity, the error bars are omitted except for one data point to indicate the typical standard errors. The mean values and standard errors (shown as red dots) of all 104 points are $C_{\cx} = -1.6 \pm 2.7,\ C_{\sx} = 0.4 \pm 2.7,\ S_{\cx} = -0.6 \pm 2.6,\ S_{\sx} = 2.0 \pm 2.6,\ C_{0} = 0.2 \pm 1.9$, and $S_{0} = -0.2 \pm 1.9$ (all values $\times 10^{-15}$).}
\end{figure*}

{\it Experiment.}---Our ring cavity is constructed from three mirrors rigidly fixed on a spacer made of Super Invar. A silicon piece of length $2\unit{cm}$ is placed along one side of the triangle and is fixed on the spacer. This silicon piece is antireflection coated and the incident angle to this piece is set to about $10^{\circ}$ to avoid the cross coupling between the counterpropagating beams. Silicon was chosen as the dielectric element because it has high transmittance and a large refractive index (measured value $n=3.69$) at $\lambda = 1550 \unit{nm}$. We obtain approximately 3 times better sensitivity compared with that obtained using optical glass as the dielectric element [see \shiki{eqM}]. The round-trip length of our cavity is $14\unit{cm}$, and the finesse is about 120 for $p$-polarized light, with the silicon piece inside the cavity. \par

We used a double-pass configuration~\cite{doublepass} for measuring the resonant frequency difference between the two counterpropagating directions. Our experimental setup is shown in \zu{opticalsetup}. We use a single-frequency laser source with a wavelength of $1550\unit{nm}$. The laser beam is fed into the ring cavity in the counterclockwise direction via a polarization maintaining fiber. The incident beam power to the cavity is about $1\unit{mW}$. The frequency of the laser beam is stabilized to the counterclockwise resonance using a piezoelectric actuator attached on the laser cavity. We used the H\"{a}nsch-Couillaud method (or polarization spectroscopy)~\cite{Hansch,Moriwaki} to obtain the error signal for the laser frequency servo. \par

The transmitted light of the counterclockwise beam is then reflected back into the cavity in the clockwise direction by a reflection mirror (labeled RM in \zu{opticalsetup}). We obtain the second error signal, which is proportional to the resonant frequency difference (from the PDs labeled PD2s and PD2p), and in this signal we search for the Lorentz violation. To obtain the second error signal, we again used the H\"{a}nsch-Couillaud method. This double-pass configuration enables a null measurement of the resonant frequency difference, with simpler configuration than that of Ref.~\cite{Baynes}. \par

All the optics are placed in a $30 \times 30 \times 17 \unit{cm}$ vacuum enclosure ($\sim 1\unit{kPa}$) to realize a stable operation. This enclosure is fixed on a turntable and rotated. Positive and reverse rotations of $420^{\circ}$ are repeated alternately with a rotational speed of $\omega_{\rm{rot}}=30^{\circ}$/sec ($f_{\rm{rot}}=0.083$ Hz). Compared with using the Earth's rotation alone, such active rotation modulates the Lorentz violation signal at a higher frequency, where the noise level is generally lower than that at the sidereal frequency.\par

{\it Data analysis.}---The data were recorded with a sampling frequency of $100 \unit{Hz}$ for 104 days between August and December 2012. During the data acquisition, the ring cavity was rotated approximately $3.7 \times 10^5$ times. \par

We extracted the SME parameters from the data by a demodulation method. From \shiki{dnu/nu}, by substituting $\theta$ with $\omega_{\rot}t$, the frequency difference $\Delta \nu \equiv \nu_{+}- \nu_{-}$ can be written as
\begin{equation} \label{demod}
  \frac{\Delta \nu}{\nu} = \mathcal{C} \cos{\omega_{\rot} t}+ \mathcal{S} \sin{\omega_{\rot} t} ,
\end{equation}
The amplitudes $\mathcal{C}$ and $\mathcal{S}$ vary with Earth's sidereal frequency $\omegao$ and are given by
\begin{eqnarray}
  \mathcal{C} &=& C_{\cx} \cos{\omegao \To} + C_{\sx} \sin{\omegao \To} + C_0 ,  \label{demod2c} \\
  \mathcal{S} &=& S_{\cx} \cos{\omegao \To} + S_{\sx} \sin{\omegao \To} + S_0 .  \label{demod2s} 
\end{eqnarray}
The six sidereal modulation amplitudes are related to the SME parameters as given in \hyou{modamps}.\par

To analyze our data, we first demodulated the data with $\omega_{\rot}$, and the modulation amplitudes $\mathcal{C}$ and $\mathcal{S}$ in \shiki{demod} were determined for each rotation. We only used an interval of $360^{\circ}$ in the middle of each $420^{\circ}$ rotation where the rotational speed is constant. This is because our ring cavity is also sensitive to the changes in the rotational speed due to the Sagnac effect~\cite{Sagnac}.\par

Next, time series data of $\mathcal{C}$ and $\mathcal{S}$ were split into 1 day intervals and fitted by the least-squares method with Eqs.~(\ref{demod2c}) and (\ref{demod2s}) to determine the six sidereal modulation amplitudes for each day. The results are shown in \zu{phasespace} as pairs of quadratic amplitudes. If we take the mean values for each amplitude, we obtain standard errors of $\sim 3 \times 10^{-15}$. No deviation from zero by more than $1 \sigma$ was found. Thus, we conclude that no significant evidence for anisotropy in the speed of light in a sidereal frame can be claimed from our data. Note that this analysis is highly independent of the choice of test theory at this point.\par

To determine the SME parameters, time series data of the six sidereal modulation amplitudes are simultaneously fitted with the corresponding expressions shown in~\hyou{modamps}. We used the weighted least-squares method to fit the data, and the obtained SME parameters and their standard errors are summarized in \hyou{result}. The $1 \sigma$ uncertainties of three SME odd-parity violation parameters $\tk_{o+}^{JK}$ are $\sim 1 \times 10^{-14}$, a factor of more than 5 better than the previous best limits obtained from an even-parity MM-type experiment~\cite{Herrmann}. The $1 \sigma$ uncertainty of the scalar parameter $\tk_{\tr}$ is $0.9 \times 10^{-10}$, more than an order of magnitude better than the previous best limit obtained from another odd-parity experiment~\cite{Baynes}. \par

Although much of the cavity length fluctuation and systematic effects are canceled out by the use of counterpropagating beams~\cite{Baynes}, there are some residual effects, which mainly originate from the alignment fluctuation of the incident beam. The static noise level of the measurement is currently limited by the vibration of the turntable. \par

In order to find the sources of systematic errors, we recorded the tilt, environmental temperature, rotational speed, and incident and transmitted laser power during the rotation. The fluctuation of these signals in a sidereal period may produce a fake Lorentz violation signal and introduce a systematic offset to the SME parameters. Although the Sagnac effect will be a problem if the fluctuation of the rotational speed changes in a sidereal period, the measured fluctuation was less than $1 \unit{mrad/sec}$ and this effect is more than 4 orders of magnitude below our statistical error. A major cause of the systematic offset was the tilt of the base of the turntable. The tilt of the fiber collimator introduced a slight change in the alignment of the incident beam. However, the measured tilt stayed within $0.2 \unit{mrad}$, and this effect was less than $10\%$ of the statistical error.\par
Another systematic error comes from uncertainty in the calibration. There was a slight drift in the calibration factor for the Lorentz violation signal, which originated from slight detuning in the laser frequency servo. This detuning was introduced by the polarization drift of the incident beam. The detuning can be monitored from the offset level of the acquired data and this calibration uncertainty was estimated to be $3\%$.\par

These systematic effects and noises will be reduced by installing a polarization control system and a reasonable vibration isolation system. Fixing all the optics monolithically on a single optical bench would also reduce the alignment fluctuation and the polarization drift of the incident beam.\par

\begin{table}
    \caption{\label{modamps} Relation between photonic SME parameters and sidereal modulation amplitudes according to Eqs.~(\ref{demod2c}) and (\ref{demod2s}). $\Omega_{\oplus}$ is the angular frequency of the Earth's orbital motion, $\chi=54.3^{\circ}$ is the colatitude of the laboratory in Tokyo and $\eta=23.4^{\circ}$ is the axial tilt of the Earth relative to the SCCEF $Z$ axis. $\To=0$ is set to the instant when the laboratory frame $y$ axis is oriented at $90^{\circ}$ right ascension.}
\begin{ruledtabular}
\begin{tabular}{cc}
       & SME amplitude \\
      \hline
$C_{\cx}$ & $-2M\cos{\chi}(\tk_{o+}^{YZ} + 2 \betao \tk_{\tr} \sin \Omegao \To)$ \\
$C_{\sx}$ & $2M\cos{\chi}(\tk_{o+}^{XZ} + 2 \betao \tk_{\tr} \cos{\eta} \cos \Omegao \To)$ \\
$C_0$ & $2M\sin{\chi}(\tk_{o+}^{XY} - 2 \betao \tk_{\tr} \sin{\eta} \cos \Omegao \To)$ \\
$S_{\cx}$ & $C_{\sx}/\cos{\chi}$ \\
$S_{\sx}$ & $-C_{\cx}/\cos{\chi}$ \\
$S_0$ & $0$ \\
\end{tabular}
\end{ruledtabular}
\end{table}

\begin{table}
\caption{\label{result} Photonic SME parameters with $1\sigma$ errors determined from this work. For comparison, the previous best limits set by \refs{Herrmann, Baynes} are also given.}
\begin{ruledtabular}
\begin{tabular}{lcc}
 & This work & Previous best \\
\hline
$\tk_{o+}^{YZ}/10^{-14}$ & $0.5 \pm 1.0$ & $-3.4 \pm 6.1$ \cite{Herrmann} \\
$\tk_{o+}^{XZ}/10^{-14}$ & $-0.6 \pm 1.2$ & $-4.5 \pm 6.2$ \cite{Herrmann} \\
$\tk_{o+}^{XY}/10^{-14}$ & $0.7 \pm 1.0$ & $-1.4 \pm 7.8$ \cite{Herrmann} \\
$\tk_{\tr}/10^{-10}$ & $-0.4 \pm 0.9$ & $3 \pm 11$ \cite{Baynes} \\
\end{tabular}
\end{ruledtabular}
\end{table}

{\it Conclusion.}---We searched for the Lorentz violation in electrodynamics using an asymmetric optical ring cavity by making use of a double-pass configuration. No clear evidence for odd-parity anisotropy in the speed of light was found at the level of $\delta c /c \lesssim 10^{-14}$. Our new constraints are the tightest directly obtained constraints to the best of our knowledge, and the improvement factors for the photonic SME odd-parity parameters $\tk_{o+}^{JK}$ and the scalar parameter $\tk_{\tr}$ are 5 and 12, respectively.\par
The use of counterpropagating resonances with a double-pass configuration enables a null experiment. This novel configuration reduces environmentally induced noise by its high common mode rejection ratio and eliminates most of the systematic effects. By sufficiently upgrading the system using the techniques mentioned above, we are expecting to further improve the sensitivity by a few orders of magnitude.\par

\acknowledgements
We thank Norikatsu Mio and Shigemi Otsuka for valuable discussions. This work was supported by JSPS Grant-in-Aid for Scientific Research (A) No. 22244049.

%


\begin{thebibliography}{20}%
\makeatletter
\providecommand \@ifxundefined [1]{%
 \@ifx{#1\undefined}
}%
\providecommand \@ifnum [1]{%
 \ifnum #1\expandafter \@firstoftwo
 \else \expandafter \@secondoftwo
 \fi
}%
\providecommand \@ifx [1]{%
 \ifx #1\expandafter \@firstoftwo
 \else \expandafter \@secondoftwo
 \fi
}%
\providecommand \natexlab [1]{#1}%
\providecommand \enquote  [1]{``#1''}%
\providecommand \bibnamefont  [1]{#1}%
\providecommand \bibfnamefont [1]{#1}%
\providecommand \citenamefont [1]{#1}%
\providecommand \href@noop [0]{\@secondoftwo}%
\providecommand \href [0]{\begingroup \@sanitize@url \@href}%
\providecommand \@href[1]{\@@startlink{#1}\@@href}%
\providecommand \@@href[1]{\endgroup#1\@@endlink}%
\providecommand \@sanitize@url [0]{\catcode `\\12\catcode `\$12\catcode
  `\&12\catcode `\#12\catcode `\^12\catcode `\_12\catcode `\%12\relax}%
\providecommand \@@startlink[1]{}%
\providecommand \@@endlink[0]{}%
\providecommand \url  [0]{\begingroup\@sanitize@url \@url }%
\providecommand \@url [1]{\endgroup\@href {#1}{\urlprefix }}%
\providecommand \urlprefix  [0]{URL }%
\providecommand \Eprint [0]{\href }%
\providecommand \doibase [0]{http://dx.doi.org/}%
\providecommand \selectlanguage [0]{\@gobble}%
\providecommand \bibinfo  [0]{\@secondoftwo}%
\providecommand \bibfield  [0]{\@secondoftwo}%
\providecommand \translation [1]{[#1]}%
\providecommand \BibitemOpen [0]{}%
\providecommand \bibitemStop [0]{}%
\providecommand \bibitemNoStop [0]{.\EOS\space}%
\providecommand \EOS [0]{\spacefactor3000\relax}%
\providecommand \BibitemShut  [1]{\csname bibitem#1\endcsname}%
\let\auto@bib@innerbib\@empty
\bibitem [{\citenamefont {Einstein}(1905)}]{Einstein}%
  \BibitemOpen
  \bibfield  {author} {\bibinfo {author} {\bibfnamefont {A.}~\bibnamefont
  {Einstein}},\ }\href@noop {} {\bibfield  {journal} {\bibinfo  {journal} {Ann.
  Phys. (Leipzig)}\ }\textbf {\bibinfo {volume} {322}},\ \bibinfo {pages} {891}
  (\bibinfo {year} {1905})}\BibitemShut {NoStop}%
\bibitem [{\citenamefont {Mattingly}(2005)}]{Mattingly}%
  \BibitemOpen
  \bibfield  {author} {\bibinfo {author} {\bibfnamefont {D.~M.}\ \bibnamefont
  {Mattingly}},\ }\href {http://www.livingreviews.org/lrr-2005-5} {\bibfield
  {journal} {\bibinfo  {journal} {Living Rev. Relativity}\ }\textbf {\bibinfo
  {volume} {8}},\ \bibinfo {pages} {5} (\bibinfo {year} {2005})}\BibitemShut
  {NoStop}%
\bibitem [{\citenamefont {Kosteleck{\'y}}\ and\ \citenamefont
  {Samuel}(1989)}]{LorentzViolation}%
  \BibitemOpen
  \bibfield  {author} {\bibinfo {author} {\bibfnamefont {V.~A.}\ \bibnamefont
  {Kosteleck{\'y}}}\ and\ \bibinfo {author} {\bibfnamefont {S.}~\bibnamefont
  {Samuel}},\ }\href {\doibase 10.1103/PhysRevD.39.683} {\bibfield  {journal}
  {\bibinfo  {journal} {Phys. Rev. D}\ }\textbf {\bibinfo {volume} {39}},\
  \bibinfo {pages} {683} (\bibinfo {year} {1989})}\BibitemShut {NoStop}%
\bibitem [{\citenamefont {Colladay}\ and\ \citenamefont
  {Kosteleck{\'y}}(1998)}]{SME}%
  \BibitemOpen
  \bibfield  {author} {\bibinfo {author} {\bibfnamefont {D.}~\bibnamefont
  {Colladay}}\ and\ \bibinfo {author} {\bibfnamefont {V.~A.}\ \bibnamefont
  {Kosteleck{\'y}}},\ }\href {\doibase 10.1103/PhysRevD.58.116002} {\bibfield
  {journal} {\bibinfo  {journal} {Phys. Rev. D}\ }\textbf {\bibinfo {volume}
  {58}},\ \bibinfo {pages} {116002} (\bibinfo {year} {1998})}\BibitemShut
  {NoStop}%
\bibitem [{\citenamefont {Kosteleck{\'y}}\ and\ \citenamefont
  {Mewes}(2002)}]{Birefringent}%
  \BibitemOpen
  \bibfield  {author} {\bibinfo {author} {\bibfnamefont {V.~A.}\ \bibnamefont
  {Kosteleck{\'y}}}\ and\ \bibinfo {author} {\bibfnamefont {M.}~\bibnamefont
  {Mewes}},\ }\href {\doibase 10.1103/PhysRevD.66.056005} {\bibfield  {journal}
  {\bibinfo  {journal} {Phys. Rev. D}\ }\textbf {\bibinfo {volume} {66}},\
  \bibinfo {pages} {056005} (\bibinfo {year} {2002})}\BibitemShut {NoStop}%
\bibitem [{\citenamefont {Eisele}\ \emph {et~al.}(2009)\citenamefont {Eisele},
  \citenamefont {Nevsky},\ and\ \citenamefont {Schiller}}]{Eisele}%
  \BibitemOpen
  \bibfield  {author} {\bibinfo {author} {\bibfnamefont {C.}~\bibnamefont
  {Eisele}}, \bibinfo {author} {\bibfnamefont {A.~Y.}\ \bibnamefont {Nevsky}},
  \ and\ \bibinfo {author} {\bibfnamefont {S.}~\bibnamefont {Schiller}},\
  }\href {\doibase 10.1103/PhysRevLett.103.090401} {\bibfield  {journal}
  {\bibinfo  {journal} {Phys. Rev. Lett.}\ }\textbf {\bibinfo {volume} {103}},\
  \bibinfo {pages} {090401} (\bibinfo {year} {2009})}\BibitemShut {NoStop}%
\bibitem [{\citenamefont {Herrmann}\ \emph {et~al.}(2009)\citenamefont
  {Herrmann}, \citenamefont {Senger}, \citenamefont {Mohle}, \citenamefont
  {Nagel}, \citenamefont {Kovalchuk},\ and\ \citenamefont {Peters}}]{Herrmann}%
  \BibitemOpen
  \bibfield  {author} {\bibinfo {author} {\bibfnamefont {S.}~\bibnamefont
  {Herrmann}}, \bibinfo {author} {\bibfnamefont {A.}~\bibnamefont {Senger}},
  \bibinfo {author} {\bibfnamefont {K.}~\bibnamefont {Mohle}}, \bibinfo
  {author} {\bibfnamefont {M.}~\bibnamefont {Nagel}}, \bibinfo {author}
  {\bibfnamefont {E.~V.}\ \bibnamefont {Kovalchuk}}, \ and\ \bibinfo {author}
  {\bibfnamefont {A.}~\bibnamefont {Peters}},\ }\href {\doibase
  10.1103/PhysRevD.80.105011} {\bibfield  {journal} {\bibinfo  {journal} {Phys.
  Rev. D}\ }\textbf {\bibinfo {volume} {80}},\ \bibinfo {pages} {105011}
  (\bibinfo {year} {2009})}\BibitemShut {NoStop}%
\bibitem [{\citenamefont {Hohensee}\ \emph {et~al.}(2010)\citenamefont
  {Hohensee}, \citenamefont {Stanwix}, \citenamefont {Tobar}, \citenamefont
  {Parker}, \citenamefont {Phillips},\ and\ \citenamefont
  {Walsworth}}]{Hohensee}%
  \BibitemOpen
  \bibfield  {author} {\bibinfo {author} {\bibfnamefont {M.~A.}\ \bibnamefont
  {Hohensee}}, \bibinfo {author} {\bibfnamefont {P.~L.}\ \bibnamefont
  {Stanwix}}, \bibinfo {author} {\bibfnamefont {M.~E.}\ \bibnamefont {Tobar}},
  \bibinfo {author} {\bibfnamefont {S.~R.}\ \bibnamefont {Parker}}, \bibinfo
  {author} {\bibfnamefont {D.~F.}\ \bibnamefont {Phillips}}, \ and\ \bibinfo
  {author} {\bibfnamefont {R.~L.}\ \bibnamefont {Walsworth}},\ }\href {\doibase
  10.1103/PhysRevD.82.076001} {\bibfield  {journal} {\bibinfo  {journal} {Phys.
  Rev. D}\ }\textbf {\bibinfo {volume} {82}},\ \bibinfo {pages} {076001}
  (\bibinfo {year} {2010})}\BibitemShut {NoStop}%
\bibitem [{\citenamefont {Trimmer}\ \emph {et~al.}(1973)\citenamefont
  {Trimmer}, \citenamefont {Baierlein}, \citenamefont {Faller},\ and\
  \citenamefont {Hill}}]{Trimmer}%
  \BibitemOpen
  \bibfield  {author} {\bibinfo {author} {\bibfnamefont {W.~S.~N.}\
  \bibnamefont {Trimmer}}, \bibinfo {author} {\bibfnamefont {R.~F.}\
  \bibnamefont {Baierlein}}, \bibinfo {author} {\bibfnamefont {E.}~\bibnamefont
  {Faller}}, \ and\ \bibinfo {author} {\bibfnamefont {H.~A.}\ \bibnamefont
  {Hill}},\ }\href {\doibase 10.1103/PhysRevD.8.3321} {\bibfield  {journal}
  {\bibinfo  {journal} {Phys. Rev. D}\ }\textbf {\bibinfo {volume} {8}},\
  \bibinfo {pages} {3321} (\bibinfo {year} {1973})}\BibitemShut {NoStop}%
\bibitem [{\citenamefont {Tobar}\ \emph {et~al.}(2005)\citenamefont {Tobar},
  \citenamefont {Wolf}, \citenamefont {Fowler},\ and\ \citenamefont
  {Hartnett}}]{Tobar2005}%
  \BibitemOpen
  \bibfield  {author} {\bibinfo {author} {\bibfnamefont {M.~E.}\ \bibnamefont
  {Tobar}}, \bibinfo {author} {\bibfnamefont {P.}~\bibnamefont {Wolf}},
  \bibinfo {author} {\bibfnamefont {A.}~\bibnamefont {Fowler}}, \ and\ \bibinfo
  {author} {\bibfnamefont {J.~G.}\ \bibnamefont {Hartnett}},\ }\href {\doibase
  10.1103/PhysRevD.71.025004} {\bibfield  {journal} {\bibinfo  {journal} {Phys.
  Rev. D}\ }\textbf {\bibinfo {volume} {71}},\ \bibinfo {pages} {025004}
  (\bibinfo {year} {2005})}\BibitemShut {NoStop}%
\bibitem [{\citenamefont {Exirifard}()}]{Exirifard}%
  \BibitemOpen
  \bibfield  {author} {\bibinfo {author} {\bibfnamefont {Q.}~\bibnamefont
  {Exirifard}},\ }\href@noop {} {}\Eprint {http://arxiv.org/abs/1010.2057}
  {arXiv:1010.2057} \BibitemShut {NoStop}%
\bibitem [{\citenamefont {Baynes}\ \emph {et~al.}(2012)\citenamefont {Baynes},
  \citenamefont {Tobar},\ and\ \citenamefont {Luiten}}]{Baynes}%
  \BibitemOpen
  \bibfield  {author} {\bibinfo {author} {\bibfnamefont {F.~N.}\ \bibnamefont
  {Baynes}}, \bibinfo {author} {\bibfnamefont {M.~E.}\ \bibnamefont {Tobar}}, \
  and\ \bibinfo {author} {\bibfnamefont {A.~N.}\ \bibnamefont {Luiten}},\
  }\href {\doibase 10.1103/PhysRevLett.108.260801} {\bibfield  {journal}
  {\bibinfo  {journal} {Phys. Rev. Lett.}\ }\textbf {\bibinfo {volume} {108}},\
  \bibinfo {pages} {260801} (\bibinfo {year} {2012})}\BibitemShut {NoStop}%
\bibitem [{\citenamefont {Klinkhamer}\ and\ \citenamefont
  {Schreck}(2008)}]{Klinkhamer}%
  \BibitemOpen
  \bibfield  {author} {\bibinfo {author} {\bibfnamefont {F.~R.}\ \bibnamefont
  {Klinkhamer}}\ and\ \bibinfo {author} {\bibfnamefont {M.}~\bibnamefont
  {Schreck}},\ }\href {\doibase 10.1103/PhysRevD.78.085026} {\bibfield
  {journal} {\bibinfo  {journal} {Phys. Rev. D}\ }\textbf {\bibinfo {volume}
  {78}},\ \bibinfo {pages} {085026} (\bibinfo {year} {2008})}\BibitemShut
  {NoStop}%
\bibitem [{\citenamefont {Hohensee}\ \emph {et~al.}(2009)\citenamefont
  {Hohensee}, \citenamefont {Lehnert}, \citenamefont {Phillips},\ and\
  \citenamefont {Walsworth}}]{Hohensee2009}%
  \BibitemOpen
  \bibfield  {author} {\bibinfo {author} {\bibfnamefont {M.~A.}\ \bibnamefont
  {Hohensee}}, \bibinfo {author} {\bibfnamefont {R.}~\bibnamefont {Lehnert}},
  \bibinfo {author} {\bibfnamefont {D.~F.}\ \bibnamefont {Phillips}}, \ and\
  \bibinfo {author} {\bibfnamefont {R.~L.}\ \bibnamefont {Walsworth}},\ }\href
  {\doibase 10.1103/PhysRevLett.102.170402} {\bibfield  {journal} {\bibinfo
  {journal} {Phys. Rev. Lett.}\ }\textbf {\bibinfo {volume} {102}},\ \bibinfo
  {pages} {170402} (\bibinfo {year} {2009})}\BibitemShut {NoStop}%
\bibitem [{\citenamefont {Altschul}(2009)}]{Altschul}%
  \BibitemOpen
  \bibfield  {author} {\bibinfo {author} {\bibfnamefont {B.}~\bibnamefont
  {Altschul}},\ }\href {\doibase 10.1103/PhysRevD.80.091901} {\bibfield
  {journal} {\bibinfo  {journal} {Phys. Rev. D}\ }\textbf {\bibinfo {volume}
  {80}},\ \bibinfo {pages} {091901} (\bibinfo {year} {2009})}\BibitemShut
  {NoStop}%
\bibitem [{\citenamefont {Kosteleck\'y}\ and\ \citenamefont
  {Russell}(2011)}]{DataTables}%
  \BibitemOpen
  \bibfield  {author} {\bibinfo {author} {\bibfnamefont {V.~A.}\ \bibnamefont
  {Kosteleck\'y}}\ and\ \bibinfo {author} {\bibfnamefont {N.}~\bibnamefont
  {Russell}},\ }\href {\doibase 10.1103/RevModPhys.83.11} {\bibfield  {journal}
  {\bibinfo  {journal} {Rev. Mod. Phys.}\ }\textbf {\bibinfo {volume} {83}},\
  \bibinfo {pages} {11} (\bibinfo {year} {2011})}\BibitemShut {NoStop}%
\bibitem [{\citenamefont {Cusack}\ \emph {et~al.}(2002)\citenamefont {Cusack},
  \citenamefont {Shaddock}, \citenamefont {Slagmolen}, \citenamefont {de~Vine},
  \citenamefont {Gray},\ and\ \citenamefont {McClelland}}]{doublepass}%
  \BibitemOpen
  \bibfield  {author} {\bibinfo {author} {\bibfnamefont {B.~J.}\ \bibnamefont
  {Cusack}}, \bibinfo {author} {\bibfnamefont {D.~A.}\ \bibnamefont
  {Shaddock}}, \bibinfo {author} {\bibfnamefont {B.~J.~J.}\ \bibnamefont
  {Slagmolen}}, \bibinfo {author} {\bibfnamefont {G.}~\bibnamefont {de~Vine}},
  \bibinfo {author} {\bibfnamefont {M.~B.}\ \bibnamefont {Gray}}, \ and\
  \bibinfo {author} {\bibfnamefont {D.~E.}\ \bibnamefont {McClelland}},\ }\href
  {\doibase 10.1088/0264-9381/19/7/379} {\bibfield  {journal} {\bibinfo
  {journal} {Class. Quantum Grav.}\ }\textbf {\bibinfo {volume} {19}},\
  \bibinfo {pages} {1819} (\bibinfo {year} {2002})}\BibitemShut {NoStop}%
\bibitem [{\citenamefont {H{\"a}nsch}\ and\ \citenamefont
  {Couillaud}(1980)}]{Hansch}%
  \BibitemOpen
  \bibfield  {author} {\bibinfo {author} {\bibfnamefont {T.~W.}\ \bibnamefont
  {H{\"a}nsch}}\ and\ \bibinfo {author} {\bibfnamefont {B.}~\bibnamefont
  {Couillaud}},\ }\href {\doibase 10.1016/0030-4018(80)90069-3} {\bibfield
  {journal} {\bibinfo  {journal} {Opt. Commun.}\ }\textbf {\bibinfo {volume}
  {35}},\ \bibinfo {pages} {441} (\bibinfo {year} {1980})}\BibitemShut
  {NoStop}%
\bibitem [{\citenamefont {Moriwaki}\ \emph {et~al.}(2009)\citenamefont
  {Moriwaki}, \citenamefont {Mori}, \citenamefont {Takeno},\ and\ \citenamefont
  {Mio}}]{Moriwaki}%
  \BibitemOpen
  \bibfield  {author} {\bibinfo {author} {\bibfnamefont {S.}~\bibnamefont
  {Moriwaki}}, \bibinfo {author} {\bibfnamefont {T.}~\bibnamefont {Mori}},
  \bibinfo {author} {\bibfnamefont {K.}~\bibnamefont {Takeno}}, \ and\ \bibinfo
  {author} {\bibfnamefont {N.}~\bibnamefont {Mio}},\ }\href {\doibase
  10.1143/APEX.2.016501} {\bibfield  {journal} {\bibinfo  {journal} {Appl.
  Phys. Express}\ }\textbf {\bibinfo {volume} {2}},\ \bibinfo {pages} {016501}
  (\bibinfo {year} {2009})}\BibitemShut {NoStop}%
\bibitem [{\citenamefont {Post}(1967)}]{Sagnac}%
  \BibitemOpen
  \bibfield  {author} {\bibinfo {author} {\bibfnamefont {E.~J.}\ \bibnamefont
  {Post}},\ }\href {\doibase 10.1103/RevModPhys.39.475} {\bibfield  {journal}
  {\bibinfo  {journal} {Rev. Mod. Phys.}\ }\textbf {\bibinfo {volume} {39}},\
  \bibinfo {pages} {475} (\bibinfo {year} {1967})}\BibitemShut {NoStop}%
\end{thebibliography}%
\end{document}